\documentclass[runningheads,a4paper]{llncs}
\usepackage{amssymb}
\usepackage{array}
\usepackage{graphicx}

\begin{document}

\title{Towards a Cloud-Based Service for Maintaining and Analyzing Data About Scientific Events}
\titlerunning{Cloud-Based Service for Analyzing Scientific Data}

 \author{Andreas Behrend, Sahar Vadhdati, Christoph Lange \and Christiane Engels}
% \institute{Institute of CS III, University of Bonn,\\
 \institute{University of Bonn,
        Institute of Computer Science III,\\
 	R\"omerstra\ss e 164, D-53117 Bonn, Germany\\
 	\email{\{behrend,vahdati,engelsc\}@cs.uni-bonn.de}\\
	\email{math.semantic.web@gmail.com}}

%\affil{Martin-Luther-Universit\"at Halle-Wittenberg,
%	Institut f\"ur Informatik,\\
%	Von-Seckendorff-Platz~1, D-06099 Halle (Saale), Germany\\
%	\texttt{brass@informatik.uni-halle.de}}
%\Copyright[nd]{Stefan Brass}
%\author[S.~Brass]{STEFAN BRASS\\
%	Martin-Luther-Universit\"at Halle-Wittenberg,
%	Institut f\"ur Informatik,\\
%	Von-Seckendorff-Platz~1, D-06099 Halle (Saale), Germany\\
%	\email{brass@informatik.uni-halle.de}}
%\institute{%\vspace*{-4mm}\vrule height 0pt depth 0pt width 0pt%
	%\vbox to 0pt{\vss\hbox{\normalsize\rm
	%In: Computational Logic --- CL'2000,
	%Sixth International Conference on Rules}
	%\hbox{\normalsize\rm and Objects in Databases,
	%Springer, LNCS~1861, 2000, pages 1063-1077.}\vspace{40mm}}
	%University of Giessen,
	%Inst.~f.~Informatik,\\
	%Arndtstr.~2, D-35392 Giessen, Germany\\
	%\email{Stefan.Brass@informatik.uni-giessen.de}}

\maketitle

%\label{firstpage}

%==============================================================================
% The Abstract:
%==============================================================================

\begin{abstract}
We propose the new cloud-based service OpenResearch for managing and analyzing data about scientific events such as conferences and workshops in a persistent and reliable way. This includes data about scientific articles, participants, acceptance rates, submission numbers, impact values as well as organizational details such as program committees, chairs, fees and sponsoring. OpenResearch is a centralized repository for scientific events and supports researchers in collecting, organizing, sharing and disseminating information about scientific events in a structured way. An additional feature currently under development is the possibility to archive web pages along with the extracted semantic data in order to lift the burden of maintaining new and old conference web sites from public research institutions. However, the main advantage is that this cloud-based repository enables a comprehensive analysis of conference data. Based on extracted semantic data, it is possible to determine quality estimations, scientific communities, research trends as well the development of acceptance rates, fees and number of participants in a continuous way complemented by projections into the future. Furthermore, data about research articles can be systematically explored using a content-based analysis as well as citation linkage. All data maintained in this crowd-sourcing platform is made freely available through an open SPARQL endpoint, which allows for analytical queries in a flexible and user-defined way.
\end{abstract}


%==============================================================================
\section{Introduction}
\label{Introduction}
%==============================================================================
%
In computer science, scientific events like conferences and workshops form one of the most important ways of realizing fast publication of papers and bringing researchers together to present and discuss new results with peers. Information about such events is usually provided via web pages which are typically hosted by scientific organizations such as universities, research institutes, etc. Unfortunately, these organizations cannot ensure a reliable archive support, such that many conference pages just vanish over time. Additionally, many auxiliary services like conference management systems, mailing lists, digital libraries etc. are employed in order to organize and disseminate data about scientific events, but these specialized systems provide selected information or analysis results only. This makes the analysis of scientific events complicated as a complete information record is usually not available. For example, if a researcher searches for a conference with moderate conference fees, a high quality ranking and a continuously growing community interested in the topic 'Graph Databases', a matching event is rather difficult to find.

Therefore, scholars frequently rely on individual experience, recommendations from colleagues or informal community wisdom in order to organize their scientific communication. Often, scholars just simply search the Web or subscribe to mailing lists in order to receive potentially relevant calls for papers (CfPs). Other sources like Google Scholar Metrics\footnote{https://scholar.google.com/intl/en/scholar/metrics.html} or the British Research Excellence Framework are used for assessing the quality and topical relevance of scientific events. So, various internet sources need to be continuously consulted in order to obtain a comprehensive picture about the current research landscape. This complex analysis needs to be complemented by a bibliographic analysis in order to identify the most important research results, related papers, promising approaches as well as the search for the most active groups for a specific research topic. Despite some technical support by digital libraries and repositories on the Web, this analysis typically requires a lot of manual exploration work, e.g., by navigating the citations provided in sets of research articles.

As a solution to these organizational problems, we proposed the cloud-based service OpenResearch\footnote{http://openresearch.org/} for persistently storing, analyzing and maintaining data about scientific events in\cite{VAAL16}. OpenResearch manages data about scientific articles, participants, acceptance rates, submission numbers, impact values as well as organizational details such as the structure of program committee, chairs, fees and sponsoring. It is a platform for automating or crowd-sourcing the collection and integration of semantically structured metadata about scholarly events.

In this paper, we introduce a design view for entities in OpenResearch complementing the existing data view. For conferences, this means the possibility to store and view conference homepages along with their factual datasheet. Conference homepages are a good source of metadata that can be used to provide a comprehensive analysis about scientific events or any other scholarly communication metadata. Storing plain HTML conference homepages in OpenResearch makes them accessible for future data analysis, as in future smarter information extraction tools might be developed that would extract from these archived homepages structured data, which would then be fed into OpenResearch's database. In fact, the benefit of the proposed solution is twofold: Having a centralized repository for scientific events supports researchers in collecting, organizing, sharing and disseminating information about scientific events in a structured way. At the same time, the burden of maintaining conference web sites is lifted from public institutions allowing to considerably lower their running cost for providing hosting services.

OpenResearch is implemented using Semantic Media Wiki\footnote{http://www.semantic-mediawiki.org/} (SMW) and provides a user interface for creating and editing semantically structured project descriptions, researcher and event profiles in a collaborative, agile 'wiki' way. By the application of SMW, semantic web features are integrated into the MediaWiki wiki engine best known from Wikipedia, enabling a comprehensive search and organization of wiki content. In particular, OpenResearch ...
\begin{enumerate}
 \item represents a central storage for information about scientific events, projects and scholars e.g., hosting and archiving conference homepages
 \item allows for freely annotating the data (e.g., quality measures, related events) based on results from user-defined queries
\item supports users in finding relevant events and event organizers in promoting their events
 \item enables researchers to analyze metadata about scientific events or any other scholarly communication metadata
\end{enumerate}
With the new functionality of providing a design view of the data, more text analysis tools need to be integrated in order to leverage the above mentioned features for data analyzes. In the future, the system will be extended by the functionality to store accepted papers of a given scientific event (e.g., by supporting the semantically structured subsets of HTML such as Research Articles in Simplified HTML (RASH)\footnote{https://github.com/essepuntato/rash} or the format supported by the dokieli editor) as well as the possibility to store and display entire conference web pages along with their semantically enriched wikis. This way, we hope that a considerable part of research communication becomes freely available and analyzable in a sustainable way.

Section 2 of this paper presents the state of the art of existing approaches and services. Section 3 reviews the background of OpenResearch and the main challenges this project addresses. In Section 4 the approach and architecture of OpenResearch is presented and Section 5 presents the services that OpenResearch provides to its end users. Section 6 presents the current status of the OpenResearch project and discusses its future perspectives. Afterwards, we summarize and conclude the paper with Section 7.

%==============================================================================
\section{Related Work}
\label{Related Work}
%==============================================================================
%
\emph{CfP classification and analysis:} CfP Manager~\cite{IT15} is an information extraction tool for CfPs from the computer science domain. It extracts metadata of events from the unstructured text of CfPs focusing on keyword detection. Because of the different representations and terminologies across research communities, this approach requires domain specific implementations. CfP Manager neither supports comprehensive data analysis nor data curation workflows involving multiple stakeholders. Hurtado Martin et al. proposed an approach based on user profiles, which takes a scholar's recent publication list and recommends related CfPs using content analysis~\cite{H13}. Xia et al. presented a classification method to filter CfPs by social tagging\cite{X10}. Wang et al. proposed another approach to classify CfPs by implementing three different methods but focus on comparing the classification methods rather than services to improve scientific communication~\cite{WW15}. Paperleap\footnote{http://www.paperleap.com/ } is a commercial tool for analyzing CfPs and provide recommendations for users. The two well-known digital libraries in the field of computer science provide event support for researchers: ACM\footnote{http://www.acm.org/conferences} hosts webpages for their conferences and IEEE upcoming events with deadlines9

\emph{CfP dissemination:} DBWorld\footnote{https://research.cs.wisc.edu/dbworld/} collects data about upcoming events and other announcements in the field of databases and information systems. WikiCFP\footnote{http://www.wikicfp.com/} is a popular service for publishing CfPs on the Web. Like DBWorld, WikiCFP only supports a limited set of structured event metadata (title, dates, deadlines), which results in limited search and exploration functionality. WikiCFP employs crawlers to track high-profile conferences. Although WikiCFP claims to be a semantic wiki, there is neither collaborative authoring nor visioning and the data is neither downloadable as RDF nor accessible via a SPARQL endpoint. CFPList\footnote{https://www.cfplist.com/} works similarly to WikiCFP but focuses on social science related subjects. Data is contributed by the community using an online form. SemanticScholar\footnote{https://www.semanticscholar.org/} offers a keyword-based search facility that shows metadata about publications and authors. It uses artificial intelligence methods for retrieving highly relevant search results but offers no arbitrary user-defined queries.

\emph{Quality Estimation Approaches:} Google Scholar Metrics (GSM) provides ranked lists of conferences and journals by scientific field based on a 5-year impact analysis over the Google Scholar citation data. 20 top-ranked conferences and journals are shown for each (sub-)field. The ranking is based on the two metrics h5-index and h5-median. GSM's ranking method only considers the number of citations, whereas we intend to offer a multi-disciplinary service with a flexible search mechanism based on several quality metrics.

\emph{Data Collections:} DBLP, one of the most widely known bibliographic databases in computer science, provides information mainly about publications but also considers related entities such as authors, editors, conference proceedings and journals. Events, deadlines and subjects are out of DBLP's scope. DBLP allows event organizers to upload XML data with bibliographic data for ingestion. The dataset of DBLP is available as an RDF dump. ScholarlyData provides RDF dumps for scientific events. A new conference ontology developed for ScholarlyData improves over already existing ones about scientific events such as the Semantic Web Dog Food ontology. Springer LOD\footnote{http://lod.springer.com/ } is a portal for publishing conference metadata collected from the traditional publishing process of Springer as Linked Open Data. All these conferences are related to Computer Science. The data is available through a SPARQL endpoint, which makes it possible to search or browse the data. A graph visualization of the results is also available. For each conference, there is information about its acronym, location and time, and a link to the conferences series. The aim of this service is to enrich Springer's own metadata and link them to related datasets in the LOD Cloud.

\emph{Other Services:} Conference.city\footnote{http://conference.city/} is a new service initialized in 2016 that lists upcoming conferences by location. For each conference, title, date, deadline, location and number of views of its conference.city page are shown. The service collects data mainly from event homepages and from mailing lists. In addition, it allows users to add a conference using a form. PapersInvited18 focuses on collecting CfPs from event organizers and attracting potential participants who already have access to the ProQuest service. ProQuest acts as a hub between repositories holding rich and diverse scholarly data. The collected data is not made available to the public. EasyChair is a conference management system allowing organizers to publish CfPs and authors find calls relevant to them. ResearchGate21 is a social networking site for sharing papers and find potential collaborators. Lanyrd\footnote{https://lanyrd.com/} is a general service that allows everyone to add or discover new professional as well as scientific events. However, the keyword-based search offered by this service hardly allows for a comprehensive data analysis and the metadata shown for an event is restricted to dates and event locations.

\emph{Conclusion:} Only Cfplist supports an advanced query search but no free access to their data by means of a LOD set. Free LOD dumps are provided by CiteSeer. ScholarlyData and Springer LOD but no advanced query support. The possibility to advertise events by means of distributing CfPs in a goal-directed way and hosting entire web pages in addition to semantic wikis is not supported at all so far.

%==============================================================================
\section{Open Research Methodology}
\label{Open Research Methodology}
\begin{figure}[t]
\center
\includegraphics[width=\textwidth]{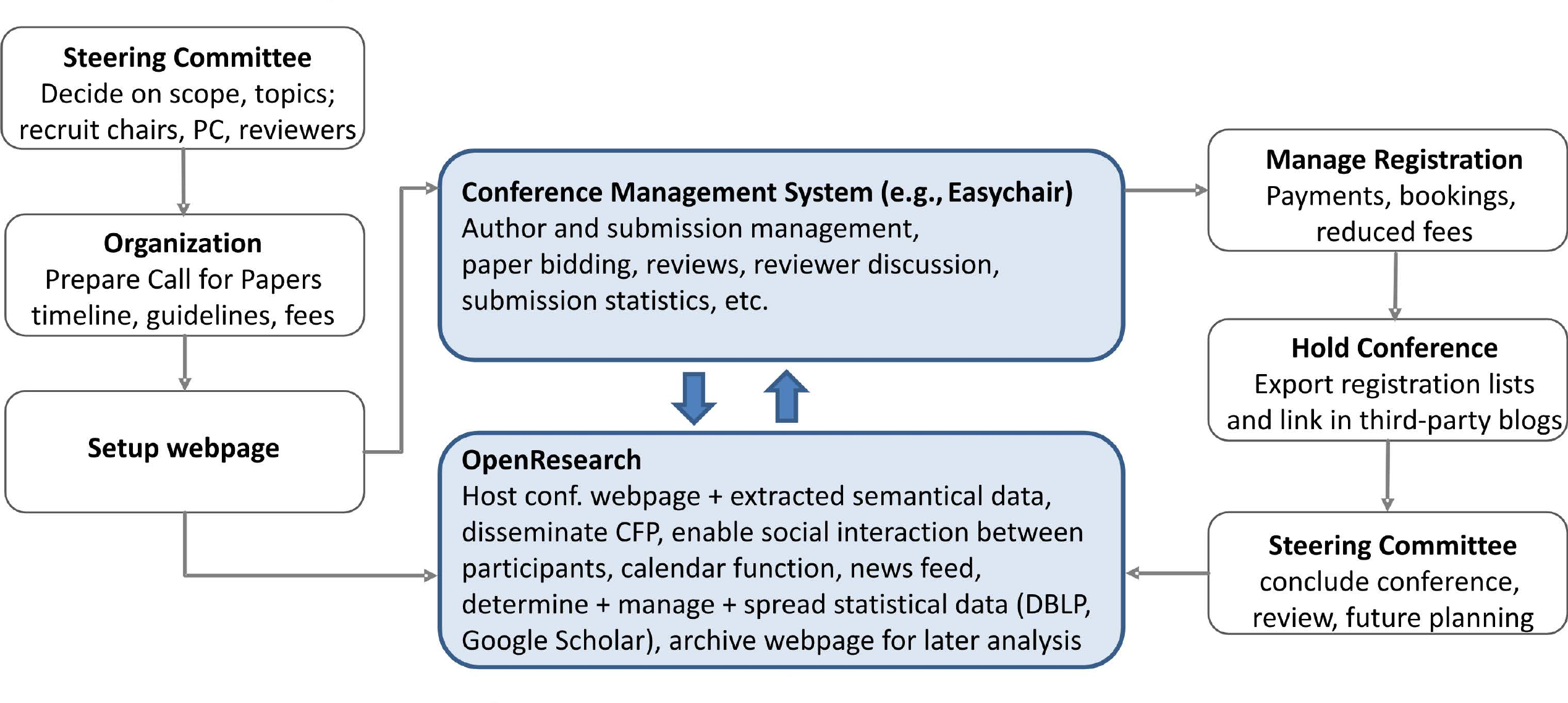}%
\caption{Role of OpenResearch in the workflow of academic conference management}
\label{fig:1}
\end{figure}

%==============================================================================
%
In~\cite{VAAL16}, we introduced OpenResearch and its general architecture and different views of quality-related queries. In this paper we extend the application range of OpenResearch as a crowd-sourcing platform which provides hosting and archiving services for conference homepages and researcher profiles. Figure~\ref{fig:1} depicts the general phases which typically occur during the organization of a conference. With the new extensions, OpenResearch can support conference organizers in various stages with a wide range of features. For realizing these features, the following problems and challenges needed to be solved.

\emph{Promoting Events:} Typically, research communities use CfPs in order to promote a scientific event. To this end, various communication channels are used to distribute CfPs like mailing lists or CfP wikis (e.g., WikiCFP). In most cases, CfPs contain unstructured data and the way of distributing them hardly allows for any systematic analysis. Consequently, potential authors or participants easily overlook relevant events and interesting analysis results like changes in topic coverage, fees and PC structure have to be manually explored. Therefore, OpenResearch provides a central repository for CfPs along with extracted semantically structured data for querying them. To this end, we developed assistants that facilitate individual as well as bulk input of CfP related data in a consistent way. Based on that, relationships between conferences and subscription lists for users can be provided in order to simplify the search for relevant events. The relevance and intrinsic complexity of this task is underlined by the fact that the recently founded company Paperleap is solely specialized on this kind of analysis alone.

\emph{Diversity of Event Structure:} There are structural differences across events, for example, events with many co-located events or sub-events, or new events emerged from multiple smaller ones. One example for the latter is the Conference on Intelligent Computer Mathematics (CICM), which results from the convergence of four conferences that used to be separate but now are tracks of a single conference. In order to support researchers in finding a suitable event, we need to automatically track and present this kind of structural information and developments for events over time. In addition, this information ought to become queryable as it may indicate changes in the underlying community. For example, in the late 90's several logic programming conferences merged into symposiums indicating a decline in the interest from the respective research community.

\emph{Tracking Event History:} A large number of events are one-off workshops which may be collocated with other events. On the other hand, most scientific events occur in series or are associated with an event series, whose individual editions took place in different locations with narrow topical changes. Researchers often need to explore several resources to obtain an overview of the previous editions of an event series to be able to estimate the quality of the next upcoming event in this series. In OpenResearch, we want to provide a quick overview of past editions and changes over time in order to support researchers in this search. Providing a complete history of event related data also allows for projections into future developments such as likely increases of conference fees or number of participants.

\emph{Addressing Different Stakeholders:} Event organizers aim to attract as many submitters as possible to their events. Publishers want to know whether they should accept a particular event's proceedings in their renowned proceedings series. Potential PC members want to decide whether it is worth spending time in the reviewing process of an event. Similarly, sponsors and invited speakers need to decide whether a certain event is worth sponsoring or attending. Researchers receiving CfP emails have to distinguish whether the event is appropriate for presenting their work. Researchers searching for events through various communication channels assess events based on criteria such as thematic relevance, feasibility of the deadline, location, low registration fee etc. The organizers of smaller events who plan to organize their event as a sub-event of a bigger event have to decide whether this is the right venue to co-locate with. These examples (for further background see~\cite{BBEK14}) prove the importance of filtering events by topic and quality from the point of view of different stakeholders. Although, quality of events have huge aspect in the competition of events, the space of information around scientific events is organized in a cumbersome way. Thus preventing events' stakeholders from making informed decisions, and preventing a competition of events around quality, economy and efficiency.

\begin{figure}[t]
\center
\includegraphics[width=\textwidth]{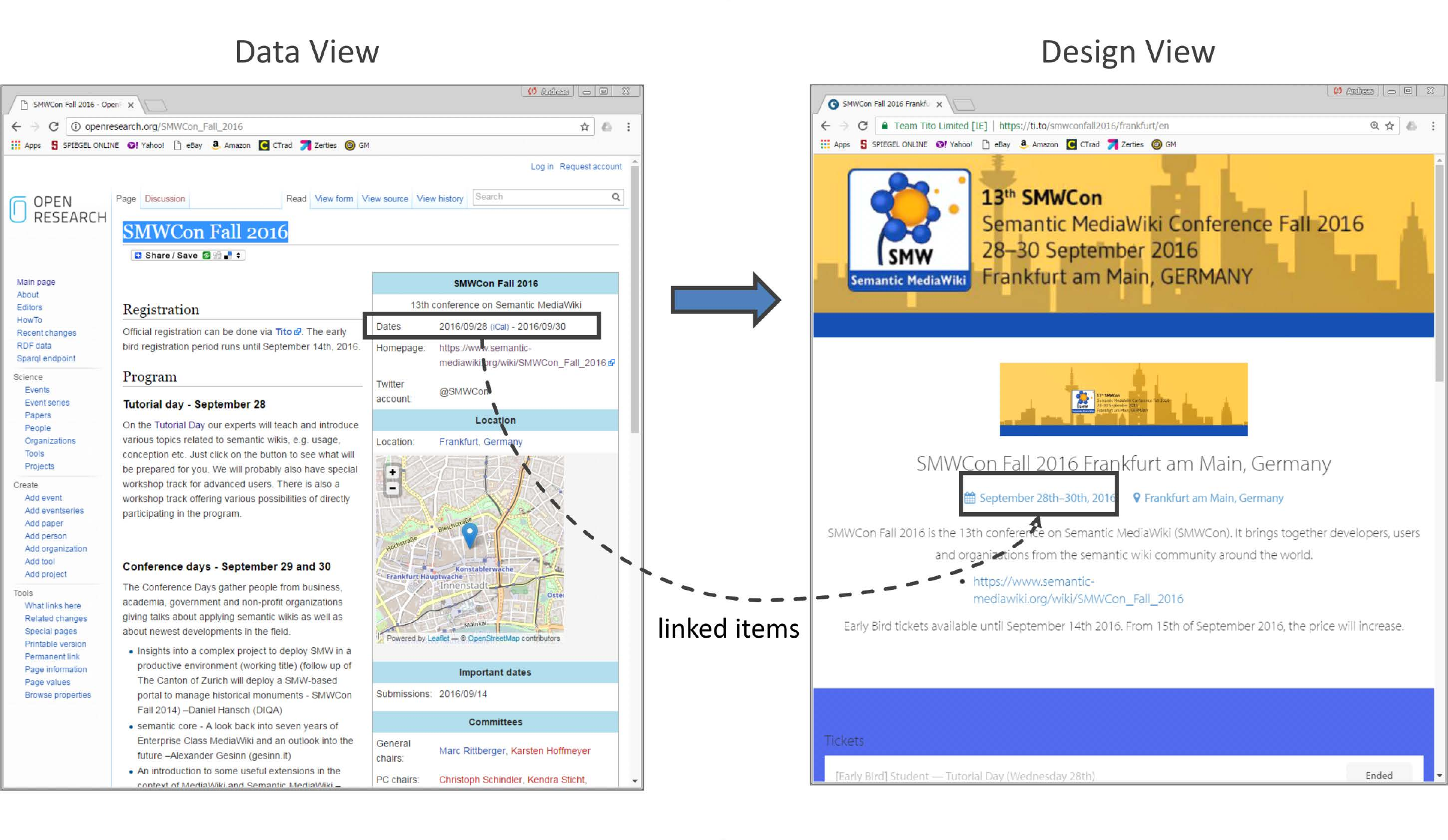}%
\caption{Two presentation views for SMWCon 2016 conference in OpenResearch}
\label{fig:2}
\end{figure}

\emph{Different Presentation Views:} Till now, OpenResearch presents event related information just in the form of wiki pages. The disadvantage of this uniform presentation of data is that the user has no control over the design in which he wants to present his research profile or well-established conference series. Therefore, we are currently extending OpenResearch in order to support two views of the data: a customized presentation view and a semantic data view. Figure~\ref{fig:2} shows the two views of the workshop SMWCon 2016 both hosted at OpenResearch. Storing the web pages of an event supports event organizers in archiving their data. In case that not all event related information has been automatically extracted, the stored web pages basically represent original sources from which further information can be subsequently retrieved. In addition, the web presence of a planned or currently active conferences can be directly managed with OpenResearch and no local island solutions are needed. Quite similar idea to GitHub Jekyll\footnote{https://github.com/jekyll/} which is a static site generator with already existed GitHub Pages, we want to establish links between extracted semantic data and its presentation in order to support the propagation of subsequent changes.

OpenResearch provides a centralized and holistic infrastructure for managing the information about scientific events and researchers. The integration of various services such as ontology based text analysis, user-defined queries and different presentation views, leads to a complex system design with a considerable number of submodules to be coordinated (cf. architecture in Section 4). OpenResearch is a cloud service with high requirements with respect to data load and access rates. The latter is not so problematic with respect to simple data access as the amount of event related data is manageable small. In order to provide high access rates for complex query results, however, we need to employ and configure a scaling SPARQL engine with a balanced resource distribution for users. At the moment, we are still using Blazegraph but performance studies (e.g, \cite{MLAN11}) already suggest system changes in the future. Finally, a reliable backup service needs to be provided in order to satisfy the quality needs of event organizers and users.

%==============================================================================
\section{Extended System Architecture}
\label{Extended System Architecture}
\begin{figure}[t]
\center
\includegraphics[width=\textwidth]{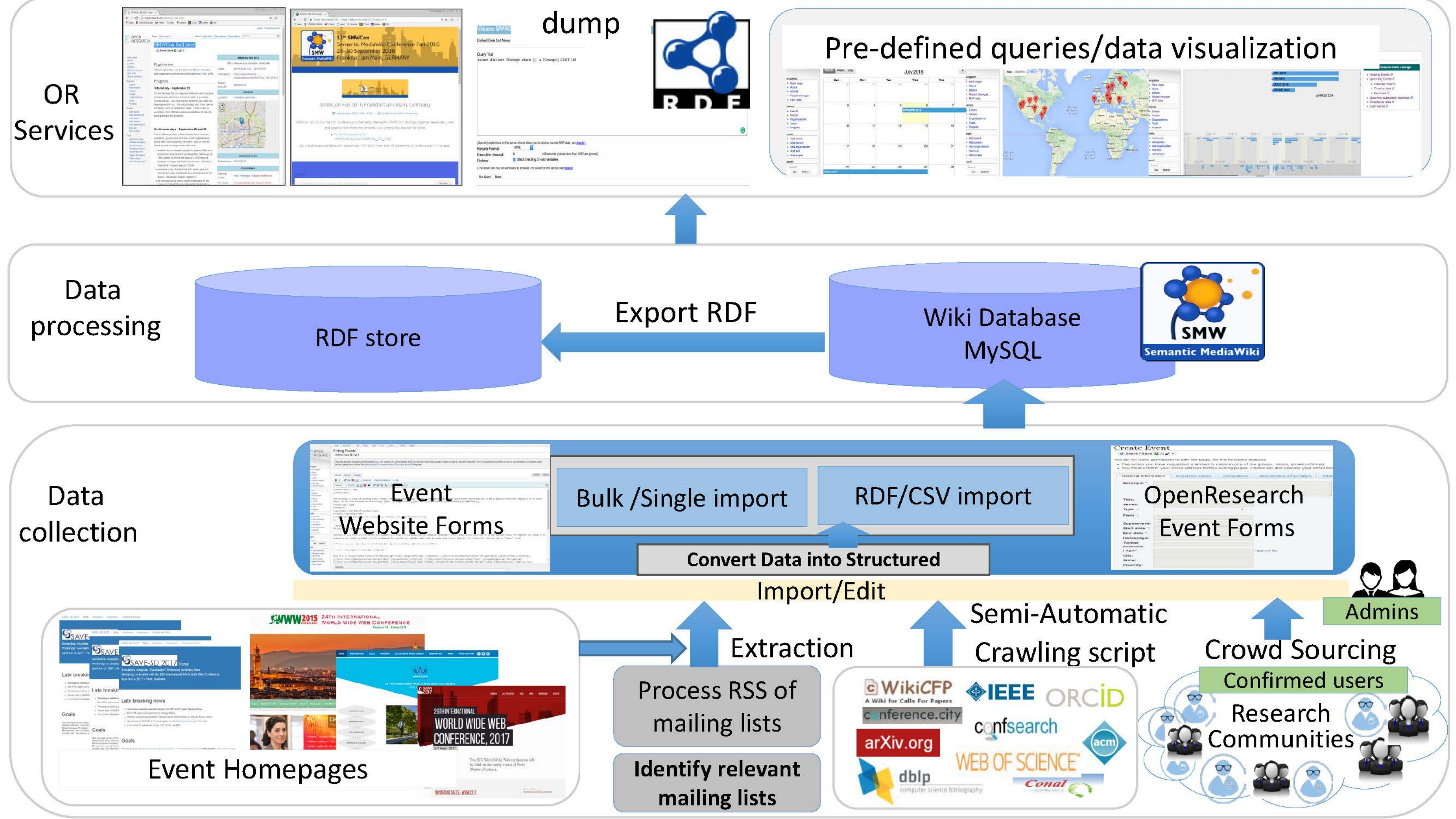}%
\caption{OpenResearch Architecture}
\label{fig:3}
\end{figure}

%==============================================================================
%
OpenResearch is a semantic wiki that balances manual (or crowd-sourced) contributions and (semi)auto-generated content. It follows the semantic wiki paradigm of instant gratification in that it aims to reduce the pain of manual data input by assistants and gives immediate access to visual analyses generated from the data just entered~\cite{AA05}. In Figure~\ref{fig:3} the three layered architecture of the system is depicted. The data gathering layer contains functionality to import event related information. To this end, data extraction tools for analyzing event homepages, web feeds as well as a crawling script for exploring various event related web sources such as WikiCFP or DBLP can be employed. In addition, registered users can create and modify event related content. OpenResearch offers flexible forms for events as well as event related web sites in order to enter data.

In the data processing layer, two database management systems are used to store and query the schema information and generated semantic triples, respectively. The underlying SMW supports various triple stores; we decided to use Blazegraph because of its performance and quality. A MySQL database instance is employed to store templates, properties and forms. OpenResearch applies a comprehensive ontology by reusing existing vocabularies (including the Semantic Web Conference Ontology and the Semantic Web Portal Ontology) and extending them with new properties25. Our ontology is stored and maintained by SMW as properties and categories. The alignment of wiki properties to ontology properties is realized using the SMW mechanism for importing vocabularies.

The service layer of OpenResearch offers various possibilities for consuming the stored data. The underlying SMW provides a full-text search facility and the visualization of their results. In addition, the RDF triple store can be accessed via a SPARQL endpoint or a downloadable RDF dump. Events, CfPs and user profiles are presented as individual wiki pages including a semantic representation of their metadata. Various query results are also visualized, including event locations on a map or event related deadlines using a calendar function.

Aside from wiki articles, OpenResearch is extended for supporting the presentation of the original design of event related web pages. This way, the system becomes an archive of old event web pages and its content which may have not been completely semantically extracted yet. The same service will be offered for user profiles such that researchers can better advertise their achievements in contrast to just presenting a uniform fact sheet about themselves. Aside from just archiving data for presentation, OpenResearch automatically analyzes the HTML code for extracting additional metadata about the given event and maintains this information in the respective fact sheet of the data view. In addition, a component is added which synchronizes changes made to these linked items in order to avoid any inconsistencies between the two views of the data. This way, even the content of a conference web page becomes partly queryable due to the incorporated information extraction mechanism.

%==============================================================================
\section{Use Cases}
\label{Use Cases}
%==============================================================================
%
In the following, we outline a few use cases that demonstrate how OpenResearch can support the analysis of metadata of scientific events. We imagine such services to be extended by interlinking OpenResearch data with other relevant datasets. Prefixes are eliminated from the sample queries.

Event-related: One challenging question that every researcher is interested in is a list of the most competitive, and thus, possibly, highest-quality conferences on their research topic.
{\small \begin{verbatim}
# Q1:conferences of topic "Semantic Web" ordered based on acceptance rate?
       SELECT ?event ?startDate ?country ?wikipage ?acceptanceRate WHERE {
            ?e a ?EventTypes ;
              rdfs:label ?event ;
              ...
              property:Acceptance_rate ?acceptanceRate ;
              property:Has_location_city ?city ;
              swivt:page ?wikipage.
            FILTER (DATATYPE(?endDate) != xsd:double &&
                    DATATYPE(?startDate) != xsd:double)
          } ORDER BY ASC(?acceptanceRate) LIMIT 10
          BINDINGS ?EventTypes {(smwont:ConferenceEvent)}
\end{verbatim} }
\noindent The result of this query is a list of high quality events with their corresponding acceptance rate, location and start date as well as a link to their wiki page. This can be extended to the event series level instead of having the list of individual editions of an event series. Another difficult and time consuming task for researchers, especially young ones, is to get an overview of event series of an given topic.
{\small \begin{verbatim}
# Q2: Life time of top five event series in the Semantic Web field
        SELECT (AVG(?num_year) AS ?averaged_life_long) WHERE
        { SELECT ?series (COUNT(DISTINCT ?startDate) AS ?num_year) WHERE
          {   ?series a category:Event_series, category:Semantic_Web.
              ...
              FILTER (DATATYPE(?endDate) != xsd:double &&
                DATATYPE(?startDate) != xsd:double)
          } GROUP BY ?series ORDER BY DESC(?num_year) LIMIT 5
        }
\end{verbatim} }
\noindent The result of this query is shown in a default table view as a number representing the average lifetime of corresponding event series in the Semantic Web field, which is 9.5 years. Identifying scientific topic movement is challenging because it is hard to define an objective notion of it~\cite{VBSL14}.

One of the possible ways to get insights on this challenge is to look at the movement of submission numbers for specific topics in their corresponding scientific events. In this way, one can see whether a topic is growing by the interest of active researchers in it. The following query addresses this issue.\pagebreak

{\small \begin{verbatim}
# Q3: Topic movement?
        SELECT  ?series ?numEvents ?topic ?events ?years WHERE{
          ?series a ?topic.
          ?topic rdfs:subClassOf category:Computer_Science.
          FILTER(?numEvents = 10).
          {
            SELECT  ?series
              (COUNT(?e) AS ?numEvents)
              (GROUP_CONCAT(DISTINCT ?e; separator="; ") AS ?events)
              (GROUP_CONCAT(DISTINCT ?startDate; separator="; ") AS ?years)
            WHERE {
              ?e rdfs:label ?event ;
                 a smwont:ConferenceEvent ;
                 property:Event_in_series ?series ;
                 property:Submitted_papers ?num_papers .
              ...
              FILTER (DATATYPE(?endDate) != xsd:double &&
                      DATATYPE(?startDate) != xsd:double).
              FILTER (?startDate >= "2007-01-01"^^xsd:date &&
                      ?endDate < "2017-01-01"^^xsd:date).
            }GROUP BY ?series
          }
\end{verbatim} }
\emph{Person-related:} In what event did a given person have some role (e.g., chair, PC member or participant). Consider Harith Alani as an example and look for his roles in different events over the last few years. Note that right now these queries are run over OpenResearch's data only; therefore other roles held by this person might not be in this list.
{\small \begin{verbatim}
# Q4: Researchers's roles in event?
  SELECT ?event ?person ?hasRole WHERE {
    ?e rdfs:label ?event ;
       ?hasRole ?person.
    ?hasRole rdfs:subPropertyOf property:Has_person.
    ?person rdfs:label "Harith Alani".
    ...
  }
\end{verbatim} }
This way a list of events and roles of the given person is generated. One can add more properties of these events to be shown in the final results. This kind of questions could help researchers who want to attend events to plan their networking when they are interested to meet certain key people in their research field.

\emph{History-related:} Recommendations are another service that OpenResearch is aiming to offer. Providing such recommendation services need to have a history of properties and values of them. Basically any property of an event can be used to obtain insights on the upcoming events by extrapolation from the history of that property. Based on the following query one could extrapolate the possible registration fee of the coming edition:\pagebreak
{\small \begin{verbatim}
# Q5: registration fees of recent events
SELECT  ?series
	   (COUNT(?e) AS ?numEvents)
	   (GROUP_CONCAT(DISTINCT ?attendFee; separator="; ") AS ?fees)
	   (GROUP_CONCAT(DISTINCT ?e; separator="; ") AS ?events)
	   (GROUP_CONCAT(DISTINCT ?startDate; separator="; ") AS ?years)
WHERE {
  		...
  		?e property:Attendance_fee ?attendFee.
  		FILTER (DATATYPE(?endDate) != xsd:double &&
                DATATYPE(?startDate) != xsd:double).
  		FILTER (?startDate >= "2014-01-01"^^xsd:date &&
                  ?endDate < "2017-01-01"^^xsd:date).
} GROUP BY ?series
\end{verbatim} }
The result of this query is a list of individual events with their registration fees. Usually, researchers start searching for possible events to submit their research results as soon as they have made interesting and new achievements. This timing might not be aligned with the schedule of relevant and high quality events. CfPs are typically announced a few months before an event, but researchers might overlook them. A good way of supporting researchers is to give them suggestions of time periods in which they can expect CfPs for relevant events. With the below query we explore over OpenResearch to find periods of year in which most events for a specific field or topic have been held.
{\small \begin{verbatim}
# Q6: When to submit?
        SELECT ?month (COUNT(?e) AS ?numEvents) WHERE
        {
          ?e rdfs:label ?event;
             ...
             a category:Semantic_Web.
          FILTER (DATATYPE(?endDate) != xsd:double &&
		          DATATYPE(?startDate) != xsd:double)
          FILTER (?startDate >= "2016-01-01"^^xsd:date &&
		  ?endDate < "2017-01-01"^^xsd:date).
          VALUES ?month {1 .. 12}
          FILTER ( month(?startDate) <= ?month &&
                   ?month <= month(?endDate) )
        } GROUP BY ?month
\end{verbatim} }
Further extending this query by considering the submission deadline, researchers can effectively plan their activities as well as exporting data about an event into applications such as conferenceLive26.

%==============================================================================
\section{Current Status and Perspectives}
\label{Current Status and Perspectives}
%==============================================================================
%
OpenResearch, currently in prototype state, is a crowd-sourcing platform with more than 300 registered users. The system has been developed according to the KIDS methodology for evolving applications~\cite{LBCGG12}. OpenResearch has so far been used on the OpenResearch.org site; however, as it is technically a set of plugins and configuration rules on top of MediaWiki and its SMW extension, it could in principle be installed anywhere else as well (using data imported from the OpenResearch.org site, or using other, site-specific data), but this has so far not been documented. There exists a rudimentary GitHub repository hosting the source code of a few import filters and with a few issues being tracked; so far only the core development team has used it.

The project is part of a greater research and development agenda for enabling open access to all types of scholarly communication metadata not just from a legal but also from a technical perspective. In addition to easy and open access to such a knowledge graph, it has the potential to serve as a unique integrated environment for comprehensive analyzes on top of the metadata. The size of the community potentially interested in collaboratively analyzing scholarly communication metadata using the extended OpenResearch software can be estimated from the venues that bring together researchers with related interests: the Semantics, Analytics, Visualization: Enhancing Scholarly Data (SAVE-SD) workshop at the International World Wide Web (WWW) Conference (yearly since 2015; usually involving ~35 authors), the Big Scholarly Data (BigScholar) workshop also at WWW (since 2014; ~30 authors), the Scholarly Big Data workshop at AAAI (since 2015; ~20 authors), the Semantic Publishing Challenge at the ESWC Semantic Web Conference (since 2014; ~20), as well as sub-communities of the Scientometrics journal, the OpenSym conference on open collaboration, and of several Semantic Web, Digital Libraries and Knowledge Management conferences.

In order to ensure future sustainability of OpenResearch we are pursuing a two-fold strategy: First, we are planning to establish an OpenResearch Foundation in order to govern the platform. Secondly, we are trying to improve the reusability of the OpenResearch software so that third parties can more easily install and operate it on their own and thus contribute to building a decentral, interlinked knowledge-base about research results.

%==============================================================================
\section{Conclusion}
\label{Conclusion}
%==============================================================================
%
We presented the OpenResearch platform, which supports researchers in finding the most suitable event for their publications according to various criteria. In particular, OpenResearch allows for comprehensively querying semantically structured metadata about scholarly events. Young or even unknown events can benefit from that and can become attractive for authors due to certain quality metrics despite still being poorly ranked in general (e.g. by impact). Organizers can also benefit from our this platform because the quantification of quality becomes transparent and possibilities of positively influencing quality metrics can be systematically explored. In addition, the temporal evolution of the assessment metrics can be investigated, which allows for identifying the stability of certain quality indicators such as the acceptance rate. This plays an important role for sponsors and other long-term partners. As an additional feature, we are investigating the archiving of web pages along with the semantic data extracted from them. First, this would lift the burden of maintaining new and old conference web sites from public research institutions; secondly, it would support evolution and evaluation of information extraction: once better information extraction tools become available, one could re-apply them to the archived conference web sites.

%==============================================================================
\section{Acknowledgments}
\label{Acknowledgments}
%==============================================================================
%
This work has been partially funded by the European Commission under grant agreements no. 643410 and the German Research Foundation (DFG) under grant agreement AU 340/9-1.

%==============================================================================
% References:
%==============================================================================

%==============================================================================
% End of File:
%==============================================================================

\end{document}